\documentclass[procedia]{easychair}

\usepackage{doc}
\usepackage{makeidx}
\usepackage{times}
\usepackage{graphicx}
\usepackage{epstopdf}
\usepackage{booktabs} 
\usepackage{subfigure}

\usepackage{listings} 
\usepackage{xcolor} 
\def\procediaConference{99th Conference on Topics of
  Superb Significance (COOL 2014)}

\title{A Novel Implementation of QuickHull Algorithm \\on the GPU}

\titlerunning{QuickHull Algorithm on the GPU \ldots}

\author{
    Jiayin Zhang\inst{1}
\and
    Gang Mei\inst{1}\thanks{Corresponding author: Dr. Gang Mei. \email{gang.mei@cugb.edu.cn}; \email{gangmeiphd@gmail.com}}
\and
    Nengxiong Xu\inst{1}
\and
    Kunyang Zhao\inst{1}\\
}

\institute{
  School of Engineering and Technology, China University of Geosciences, 100083, Beijing, China\\
  \email{\{zhangjy,gang.mei,xunengxiong,kunyang\}@cugb.edu.cn}
 }

\authorrunning{J. Zhang, G. Mei, N.Xu and K. Zhao}

\begin{document}

\maketitle

\keywords{GPU; Parallelization; Convex Hull; QuickHull; Divide-and-Conquer}

\begin{abstract}
We present a novel GPU-accelerated implementation of the 
QuickHull algorihtm for calculating convex hulls of planar point sets. We also 
describe a practical solution to demonstrate how to efficiently implement a 
typical Divide-and-Conquer algorithm on the GPU. We highly utilize the 
parallel primitives provided by the library \textit{Thrust} such as the parallel segmented 
scan for better efficiency and simplicity. To evaluate the performance of 
our implementation, we carry out four groups of experimental tests using two 
groups of point sets in two modes on the GPU K20c. Experimental results indicate 
that: our implementation can achieve the speedups of up to 10.98x over the 
state-of-art CPU-based convex hull implementation Qhull \cite{1qhull}. In addition, 
our implementation can find the convex hull of 20M points in about 0.2 
seconds.  
	
\end{abstract}


%
%

\section{Introduction}
\label{sec:introduction}

The calculating of the convex hull of a set of planar points is to find the 
convex polygon that encloses all the points. Several classic algorithms have 
been developed since 1970, including the Graham scan \cite{2graham1972efficient}, Gift 
wrapping \cite{3jarvis1973identification}, Incremental method \cite{4kallay1984complexity}, Divide-and-Conquer 
\cite{5preparata1977convex}, Monotone chain \cite{6andrew1979another}, and QuickHull \cite{7barber1996quickhull}. 
Several recent efforts have also been conducted to develop efficient 
convex hull algorithms \cite{8xing2014efficient,9liu2012fast}.

The finding of convex hulls of large sets of points is in general 
computationally expensive. An effective strategy for dealing with this 
problem is to calculate convex hulls in parallel. In recent years, to 
improve the computational efficiency of calculating convex hulls, several 
worthful contributions have been made to re-design and implement sequential 
convex hull algorithms in parallel by exploiting the power of massively 
computing on the GPU \cite{10srikanth2009parallelizing,11jurkiewicz2011efficient,12srungarapu2011fast,13stein2012cudahull,14tang2012gpu,15tzeng2012finding,16white2012divide,17gao2013ghull,18gao2013flip}. Most of these GPU-accelerated 
implementations are developed based upon the 
famous QuickHull algorithm \cite{7barber1996quickhull}. 

For example, Srikanth, Kothapalli, Govindarajulu and Narayanan \cite{10srikanth2009parallelizing} first 
parallelized the QuickHull algorithm to accelerate the 
calculating of 2D convex hulls; and then Srungarapu, Reddy, Kothapalli 
and Narayanan \cite{12srungarapu2011fast} improved the above work and achieved better efficiency. 
Tzeng and Owens \cite{15tzeng2012finding} presented a framework for accelerating the 
computing of convex hull in the Divide-and-Conquer fashion by taking 
advantage of QuickHull. Similarly, by also utilizing the QuickHull approach, Stein, Geva and El-Sana \cite{13stein2012cudahull} presented a novel GPU-accelerated 
implementation of 3D convex hull algorithm. 

In this paper, we present a novel GPU-accelerated implementation of the 
famous QuickHull algorithm for computing the convex hulls of planar point 
sets. We also describe a practical solution to demonstrate how to implement 
a typical Divide-and-Conquer algorithm on the GPU. Our implementation is 
quite similar to the one introduced by Tzeng and Owens \cite{15tzeng2012finding}; but there 
are several significant differences between our implementation and that of 
Tzeng and Owens. Our implementation can achieve a speedup of up to 10.98x 
over a standard sequential CPU implementation, and can find the convex hull 
a set of 20M points in about 0.2 seconds. 

The rest of the paper is organized as follows. Section \ref{sec:background} introduces several 
background concepts behind our implementation. Section \ref{sec:idea} describes the basic 
ideas behind our implementation. Section \ref{sec:implement} further presents some 
implementation details. Then Section \ref{sec:results} gives several groups of experimental 
results, while Section \ref{sec:discuss} discusses the result. Finally, Section \ref{sec:conclusion} concludes 
this work.


\section{Segment and Segmented Scan}
\label{sec:background}

\begin{figure}[!h]
	\caption{Segments and two representation forms of segments}
	\centering
	\includegraphics[scale = 0.75]{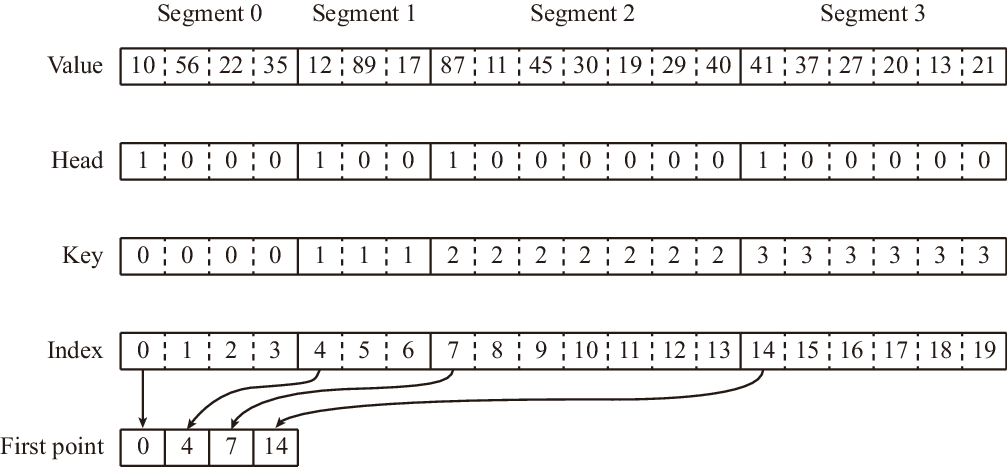}
	\label{fig:segment}
\end{figure}

Segments are contiguous partitions of the data which are maintained by 
segment flags \cite{19blelloch1990vector,20sengupta2007scan}. There are typically two forms for representing 
segments: the first one is to use a set of head flags; and the other is to use a 
set of keys; see Figure \ref{fig:segment}. A head flag marks the beginning of a segment (called 
the segment \textit{head}). A key indicates the index of the segment that each element / 
value in a given array belonging to. 

Segmented scan generalizes the scan primitive by allowing scans on arbitrary 
segments (``partitions'') of the input vector \cite{19blelloch1990vector}. To implement segmented 
scan in parallel, Sengupta, Harris, Zhang and Owens \cite{20sengupta2007scan} introduced the 
segmented scan primitive to the GPU. And currently both the libraries CUDPP \cite{21cudpp} and Thrust \cite{22bell2011thrust} both provide efficient GPU-accelerated segmented scan primitives.

Thrust is a C++ template library for CUDA based on the Standard Template 
Library (STL). Thrust allows users to implement high performance parallel 
applications with minimal programming effort through a high-level interface 
that is fully interoperable  with CUDA C \cite{23thrust}.
Thrust provides a rich collection of data structures and data parallel 
primitives such as scan, sort, and reduce, which can be composed together to 
implement complex algorithms with concise, readable source code. 

\section{Basic Ideas behind Our Implementation}
\label{sec:idea}

The most important idea behind our implementation is to directly operate the 
data in the input arrays that are originally allocated to store the 
coordinates of input points, rather than in the additionally allocated 
arrays or splitting the input data into separate arrays.

The QuickHull algorithm is a Divide-and-Conquer method, which tends to 
divide the input data set into subsets and then handles these subsets 
recursively. On the GPU, an effective strategy is to divide the input data set 
into subsets, but do not store them in separated arrays with different 
sizes. Instead, all the data of the subsets are still stored in the input 
data array, but the data of each subset is stored into a \textit{Segment} (i.e., a 
consecutive piece / partition of data) \cite{19blelloch1990vector}. Operations carried out for 
each subset is exactly the operations for each segment \cite{15tzeng2012finding}. We adopt this strategy to develop our implementation.

\begin{figure}[h!]
	\caption{Procedure of the 2D CUDA QuickHull on the GPU (without 
		preprocessing)}
	\centering
	\includegraphics[scale = 0.75]{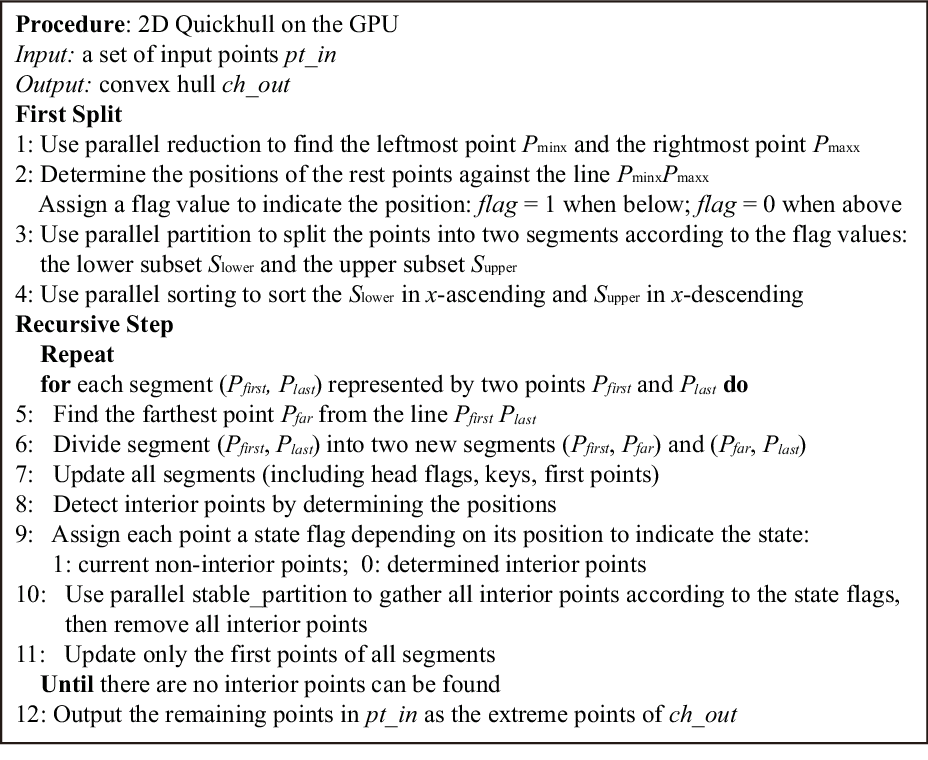}
	\label{fig:gpu;qhull}
\end{figure}

The procedure of our implementation is presented in Figure \ref{fig:gpu;qhull}. 
In our implementation, after splitting the input set of points into the 
lower and the upper subsets, we sort the two subsets separately according to 
the $x$-coordinates. After this sorting, either the lower or the upper subset 
of sorted points can be considered as a \textit{Monotone Chain} \cite{6andrew1979another}; in addition, both the 
above chains can be further considered as two halves of a general polygon. 
If we detect and remove those vertices of the general polygon that have the 
interior angles greater than 180 degrees, then 
we can obtain the desired convex hull.
The above idea of first sorting and then removing non-extreme points / 
concave vertices was introduced in  
\cite{6andrew1979another} and \cite{24melkman1987line}. 

Therefore, the basic idea behind our implementation is to 
``Find-and-Remove''. In the step of first split, we divide the input points 
and then sort them to \textbf{virtually} form a general polygon. In the 
subsequent recursive step, we recursively first \textit{find} those non-extreme points, 
and then \textit{remove} them to guarantee that all the remaining points are completely 
extreme points of the expected convex hull.


\section{Implementation Details}
\label{sec:implement}

\subsection{Data Storage and Data Layout}

We allocate several arrays on the device side to store the coordinates of 
planar points, information about segments, and other required values such as 
distances; see Table \ref{tab:data}.

\begin{table}[htbp]
	\caption{Allocated arrays for storing data on the device}
	\begin{center}
		\begin{tabular}{|l|p{280pt}|}
			\hline
			Array& 
			Usage \\
			\hline
			\texttt{float x[n]}& \textit{x} coordinates \\
			\hline
			\texttt{float y[n]}& \textit{y} coordinates \\
			\hline
			\texttt{float dist[n]}& Distances \\
			\hline
			\texttt{int head[n]}& Indicator of the first point of each segment \par (1: Head point; 0: Not a head point) \\
			\hline
			\texttt{int keys[n]}&	Index of the segment that each point belongs to \\
			\hline
			\texttt{int first{\_}pts[n]}& 	Index of the first point of each segment \\
			\hline
			\texttt{int flag[n]}& 	Indicate whether a point is an extreme point or an interior point \par (1: Potential extreme point; 0: Determined interior point) \\
			\hline
		\end{tabular}
		\label{tab:data}
	\end{center}
\end{table}

\subsection{The Preprocessing Procedure}

Before performing the QuickHull algorithm on the GPU, we first carry out a 
preprocessing procedure to filter the input points. The objective of this 
preprocessing procedure is to reduce the number of points by discarding 
those points that are not needed for consideration in the subsequent stage 
of calculating the desired convex hull. 

We use the parallel reduction to find the extreme points with min or 
max \textit{x} or \textit{y} coordinate. In more details, we adopt the
\texttt{thrust::minmax{\_}element(x.begin(), x.end())} to find the leftmost and the 
rightmost points, and similarly use the \texttt{thrust::minmax{\_}element(y.begin(), 
y.end())} to obtain the topmost and the bottommost points. These four extreme 
points are then used to form a convex quadrilateral.

We also design a simple CUDA kernel to check each point to determine whether 
it locates inside the quadrilateral. In the kernel, each thread is 
responsible for determining the position of only one point P$_{i}$, i.e., 
whether or not a point falls into the formed convex quadrilateral. If does, 
the corresponding indicator value \texttt{flag[i]} will be set to 0, otherwise, the 
value \texttt{flag[i]} is still kept as 1. 

\subsection{The First Split}

The first split of the QuichHull algorithm is to divide the set of input 
points into two subsets, i.e., the lower and the upper subsets, using the 
line segment $L$ formed by the leftmost and the rightmost points. Those points 
that locate below the $L$ are grouped into the lower subset, while the ones 
distributed above the $L$ are contained in the upper subset.

We develop another quite simple kernel to perform the above split procedure. 
In this kernel, each thread takes the responsibility to determine the 
position of only one point with respect to the line segment $L$. In this step, 
we temporarily use the values \texttt{int} \texttt{flag[n]} to indicate the positions: if the 
point P$_{i}$ locates below the $L$, in other words, if P$_{i}$ belongs to the 
lower subset, then the corresponding indicator value \texttt{flag[i]} will be set to 
1, otherwise 0. 

After determining the positions of all points, it is needed to gather the 
points belonging to the same subset such as the lower one together according 
to the indicator values \texttt{int} \texttt{flag[n]}. We realized this procedure by simply 
using the function \texttt{thrust::partition()}. Those points with the indicator 
value 1, i.e., the lower points, will be placed into the first consecutive 
half of the input array (a segment of points), while the upper points will 
be grouped into another consecutive half (another segment of points). In 
subsequent steps, operations will be performed in the segments of points. 

\subsection{The Recursive Procedure}

\subsubsection{Finding the Farthest Point}

The first step in the recursive procedure is to find the farthest points in 
each segment, which includes two remarkable issues. The first is to 
calculate the distance for all points in parallel; and the other is to find 
those points with the farthest distances for all segments in parallel.

\textbf{Calculating the Distance}. 
The calculation of the distance from a point to a line is quite 
straightforward. However, in our implementation, it is needed to calculate 
the distances from different points to different lines simultaneously in 
parallel. This calculation is not so easy to implement in practice. This is 
because that: (1) for each segment, it is needed to compute the distance 
from each of those points belonging to this segment to the line formed by 
the first points and the last point; (2) for any two segments, their first 
point and last points are different. 

In our implementation, for each point P$_{i}$, we use the value 
\texttt{first{\_}pt[i]} to record the index of the first point of that segment it 
belongs to. Since segments are stored consecutively, the first point of the 
($j$+1)$^{th}$ segment is exactly the last point of the $j^{th}$ segment except 
for the last segment. Note that the last point of the last segment is the point P$_{0}$. Therefore, it is easy to obtain the index of the last point 
for each segment, and the distance from the point P$_{i}$ to the line formed 
by the first point and the last point.

\textbf{Find the Farthest Point}. 
After calculating the distances for those points in different segments, it 
is needed to find the farthest point in each segment. The finding of the farthest points for all segments in parallel is 
not so easy. This is because there are more than two segments that are needed 
to find their farthest points. Their farthest points are private, 
segment-specific. In this case, it is unable to perform a global parallel 
reduction, but is able to employ a segmented parallel reduction to find the 
greatest distance for each segment of points. 

The segmented parallel reduction is designed in Thrust to make a parallel 
reduction for more than one segment of points. It can be used to find the 
min or max values in several segments in parallel. We employ the parallel 
primitive \texttt{thrust::reduce{\_}by{\_}key()} in our implementation to find the farthest points / maximum distances for all segments in parallel.

\subsubsection{The First Round of Updating Segments}

After finding the farthest points, each segment is then typically divided 
into two smaller sub segments using the farthest point. This means that 
the old segments are replaced with new segments. To create new segments, 
the following information of segments is needed to be updated:

(1) Head flags

The head flags of the farthest points are needed to be modified from 0 to 1, 
which means each of the farthest points becomes the first point of a new 
segment and a determined extreme points of the desired convex hull. The head flags of other points are kept unchanged.

(2) Keys

The updating of keys can be very easily realized by performing a global 
inclusive scan for the head flags. For that in each segment only the head 
flag of the first point is 1, the sum of the head flags can be considered as 
the index of the segment (i.e., the keys). Noticeably, the sum of the head 
flags is one-based rather than zero-based (i.e., starting from 1 rather 0). 
To make the indices become much easier to be used, we further modify the 
keys from one-based to zero-based by performing a parallel subtracting. 

(3) Indices of first points

After updating the head flags and keys of each segment, the corresponding 
indices of the first points are no longer valid, and thus needed to be 
updated. Before updating the indices of the first points, we first assign a 
global index for each of the remaining points, then check each point whether 
it is the first point according to the head flags. If the head flag of a 
point P$_{i}$ is 1, then this point must be a first point of a segment and 
its index is exactly $i$. 

%
%
%
%
%

\subsubsection{Discarding Interior Points}

The discarding of interior points is to first check whether or not a point 
locates inside the triangle formed by the first point of the segment 
(denoted as A), the last point of the segment (denoted as B), and the 
farthest point in this segment (denoted as C). 

Let $\bigtriangleup$ACB denote the triangle, the determining of the points' positions with 
respect to the triangle $\bigtriangleup$ACB is to check whether those points in this segment 
locate on the right side of the directed line AC and the directed line CB. If 
does, then it is not an interior point, and its corresponding indicator 
value \texttt{flag[i]} is set to 1; otherwise, it is an interior point and the value 
\texttt{flag[i]} must modified to 0. Similarly, for each point in the segment (C,B), it 
is only needed to check whether it falls on the right side the directed line CB.

\begin{figure}[h!]
	\caption{Partitioning according to flags}
	\centering
	\includegraphics[scale = 0.75]{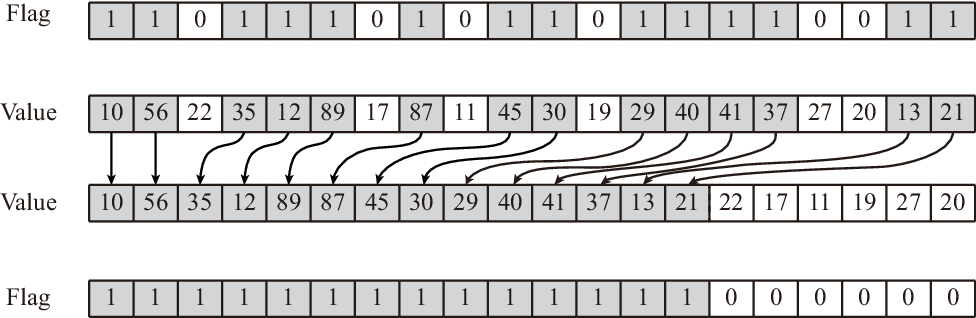}
	\label{fig:partition}
\end{figure}

After determining all interior points in this recursion, we employ a 
parallel partitioning procedure to gather all interior points together 
according to the indicator values \texttt{int flag[n]}; see a simple illustration in Figure \ref{fig:partition}. Noticeably, to maintain the 
relative order of input points, we use the function 
\texttt{thrust::stable{\_}partition()} rather than the function \texttt{thrust::partition()}. After the partitioning, we make an operation \texttt{resize()} to remove all the interior points found in this recursion. 

\subsubsection{The Second Round of Updating Segments}

This round of updating the segments is nearly the same to the first round of 
updating. The reason why this round of updating is needed to be performed is 
that: after removing some interior points, the points belonging to some 
segments are removed. 
In this case, the global indices of all the remaining point are not 
consecutive and thus need to be rearranged; see the first round of updating 
for more details about the above updating.
Noticeably, the head flags and the keys for the remaining points are still 
correct, and do not need to be updated. 
Only the index of the first point of each segment needs to be updated.

\section{Results}
\label{sec:results}

	We perform our experimental tests using the NVIDIA Tesla K20c graphics card with 4GB memory and CUDA v5.5. The CPU experiments are performed on Windows 7 SP1 with an Intel E5-2650 (2.60GHz)  and 96GB of RAM memory. 
Two groups of test data is employed in two modes to evaluate the performance of 
our implementation. The first group of test data includes 8 sets of points 
that randomly distributed in the unit square. 
The other group of test data is derived from publicly available 3D mesh 
models by projecting all vertices of each mesh model into the XY plane. 
These mesh models are directly obtained from the Stanford 3D Scanning 
Repository and the GIT Large Geometry Models Archive. 
In addition, we carry out the experimental tests in two modes:
(a) Mode 1: in this mode, the preprocessing procedure is employed, 
(b) Mode 2: in this mode, the preprocessing procedure is not used. 

We test the efficiency of our implementation using the points that are 
randomly distributed in the unit square and  the points derived from 3D mesh models, and then compare the efficiency with 
that of the Qhull library; see Tables \ref{tab:point:k20c} and \ref{tab:mesh:k20c}. The experimental results 
show that: our implementation can achieve the speedups of up to 10.98x and 
8.30x in the Mode 1 and Mode 2, respectively. In addition, it costs about 
0.2 seconds to compute the convex hull of 20M points in the best case. 

\begin{table}[h!]
	\caption{Comparison of running time (/ms) for points distributed in a square 
		on K20c}
	\begin{center}
		\begin{tabular}{|p{55pt}|p{55pt}|p{55pt}|p{55pt}|p{55pt}|p{55pt}|}
			\hline
			\raisebox{-1.50ex}[0cm][0cm]{Size}& 
			\raisebox{-1.50ex}[0cm][0cm]{Qhull}& 
			\multicolumn{2}{|c|}{Our implementation} & 
			\multicolumn{2}{|c|}{Speedup}  \\
			\cline{3-6} 
			& 
			& 
			Mode 1& 
			Mode 2& 
			Mode 1& 
			Mode 2 \\
			\hline
			100K& 
			16& 
			37.2& 
			61.7& 
			0.43& 
			0.26 \\
			\hline
			200K& 
			32& 
			36.4& 
			65.7& 
			0.88& 
			0.49 \\
			\hline
			500K& 
			62& 
			41.7& 
			69.2& 
			1.49& 
			0.90 \\
			\hline
			1M& 
			109& 
			44.8& 
			73.9& 
			2.43& 
			1.47 \\
			\hline
			2M& 
			234& 
			55.9& 
			86.2& 
			4.19& 
			2.71 \\
			\hline
			5M& 
			561& 
			77.9& 
			118.5& 
			7.20& 
			4.73 \\
			\hline
			10M& 
			1029& 
			124.7& 
			161.7& 
			8.25& 
			6.36 \\
			\hline
			20M& 
			2262& 
			206.0& 
			272.5& 
			10.98& 
			8.30 \\
			\hline
		\end{tabular}
		\label{tab:point:k20c}
	\end{center}
\end{table}

\begin{table}[h!]
	\caption{Comparison of running time (/ms) for points derived from 3D models on K20c}
	\begin{center}
		\begin{tabular}{|p{90pt}|p{30pt}|p{30pt}|p{40pt}|p{40pt}|p{40pt}|p{40pt}|}
			\hline
			\raisebox{-1.50ex}[0cm][0cm]{3D Model}& 
			\raisebox{-1.50ex}[0cm][0cm]{Size}& 
			\raisebox{-1.50ex}[0cm][0cm]{Qhull}& 
			\multicolumn{2}{|c|}{Our implementation} & 
			\multicolumn{2}{|c|}{Speedup}  \\
			\cline{4-7} 
			& 
			& 
			& 
			Mode 1& 
			Mode 2& 
			Mode 1& 
			Mode 2 \\
			\hline
			Armadillo& 
			172K& 
			16& 
			47.5& 
			61.7& 
			0.34& 
			0.26 \\
			\hline
			Angel& 
			237K& 
			47& 
			50.5& 
			62.6& 
			0.93& 
			0.75 \\
			\hline
			Skeleton Hand& 
			327K& 
			32& 
			49.5& 
			61.8& 
			0.65& 
			0.52 \\
			\hline
			Dragon& 
			437K& 
			62& 
			55.9& 
			67.4& 
			1.11& 
			0.92 \\
			\hline
			Happy Buddha& 
			543K& 
			63& 
			56.8& 
			76.7& 
			1.11& 
			0.82 \\
			\hline
			Turbine Blade& 
			882K& 
			125& 
			64.3& 
			72.3& 
			1.94& 
			1.73 \\
			\hline
			Vellum Manuscript& 
			2M& 
			219& 
			63.2& 
			91.5& 
			3.47& 
			2.39 \\
			\hline
			Asian Dragon& 
			3M& 
			359& 
			78.4& 
			129.8& 
			4.58& 
			2.77 \\
			\hline
			Thai Statue& 
			5M& 
			515& 
			84.4& 
			142.6& 
			6.10& 
			3.61 \\
			\hline
			Lucy& 
			14M& 
			1404& 
			141.3& 
			223.3& 
			9.94& 
			6.29 \\
			\hline
		\end{tabular}
		\label{tab:mesh:k20c}
	\end{center}
\end{table}


\section{Discussion}
\label{sec:discuss}

\subsection{Comparison}

The basic ideas behind our 
implementation is similar to those behind the implementation of Tzeng 
and Owens \cite{15tzeng2012finding}. The first of the same ideas is that: we perform the 
Divide-and-Conquer operations directly in the input arrays (i.e., the input 
sets of points), rather than on additionally allocated arrays. The second is 
that: the Divide-and-Conquer procedures of 
QuickHull algorithm are realized by creating, updating, or removing 
\textit{Segments}. The third similar feature is that: both of the implementations are 
developed by strongly exploiting the data-parallel primitives, parallel 
(global) scan and parallel segmented scan. 

However, we have our own ideas; and there are 
several significant differences. The first difference is that: after dividing the input set of points into the lower and the upper subsets, we further sort 
the above two subsets separately according to the \textit{x}-coordinates, and maintain 
the relative order of those sorted points unchanged in the subsequent 
procedure of recursively removing interior points. In contrast, Tzeng 
and Owens \cite{15tzeng2012finding} do not sort the subsets of points or keep the relative order 
of points. 

The second difference is the creating and removing of segments: Tzeng and 
Owens create new segments using the farthest point by permuting the points 
located in the old segment, while we also create new segments using the 
farthest point but we do not permute points. When removing segments, we at 
least retain the first point (i.e., the head point) of each segment for that 
this point is definitely an extreme point of the desired convex hull, while 
in the implementation of Tzeng and Owens a segment is probably completely 
removed. This difference is due to the different schemes of updating 
segments. 

Another difference is that: we adopt the library Thrust \cite{22bell2011thrust} 
for the use of several efficient data-parallel primitives such as parallel scan, 
segmented scan, reduction, and sorting, while in \cite{15tzeng2012finding} Tzeng and Owens 
develop their implementation by strongly exploiting the library CUDPP \cite{21cudpp}. The reason why we choose to use the library Thrust rather than 
CUDPP is that: Thrust has been integrated in CUDA toolkit, and can be much 
easier to be used in practice. 

\subsection{Performance Impact of Preprocessing}

\begin{figure}[h!]
	\centering
	\subfigure[Points distributed in unit square]{
		\label{fig:mode:k20c:a}       
		\includegraphics[scale = 0.5]{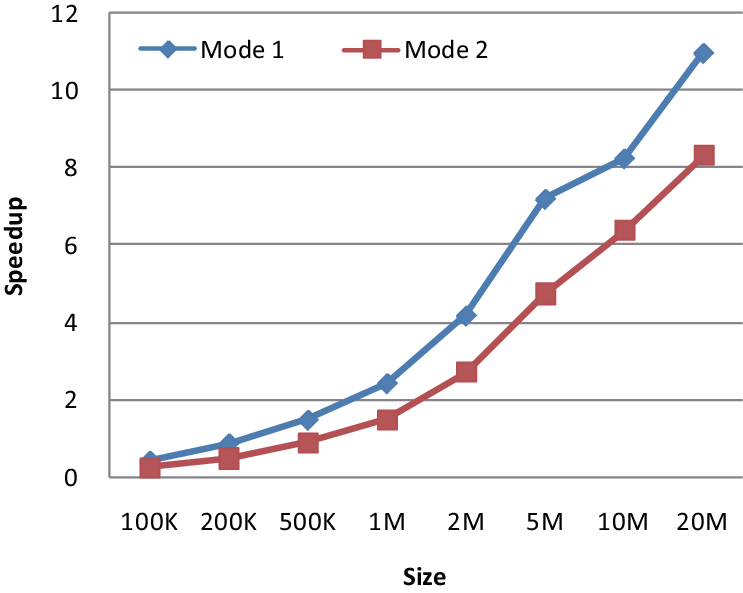}
	}
	\hspace{1em}
	\subfigure[Points derived from 3D mesh models]{
		\label{fig:mode:k20c:b}       
		\includegraphics[scale = 0.5]{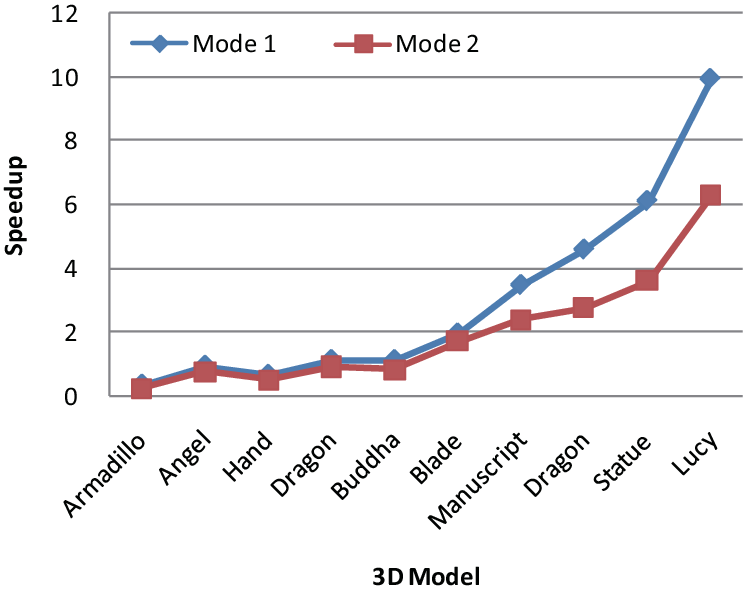}
	}
	\caption{Comparison of speedups in two modes on K20c}
	\label{fig:mode:k20c}       
\end{figure}

We have observed that: our 
implementation executing in the Mode 1 is much faster than that in Mode 2; 
see Figure \ref{fig:mode:k20c}. This behavior is probably due to the facts that: 
(1) more than 50{\%} input points can be discarded in this preprocessing; 
(2) the performance benefit from the discarding of interior points is more 
than the performance penalty lead by the discarding. 

Besides the above mentioned improvement in computational efficiency, another benefit of 
adopting the preprocessing procedure is that: the memory usage on the device 
side is much less. This is because that: after the preprocessing only less 
than 50{\%} input points remain. Therefore, in the subsequent procedure of 
computing the convex hull, it is only needed to allocate much less memory on the device side 
for the arrays such \texttt{float} \texttt{x[n]}, \texttt{y[n]}, \texttt{dist[n]} and \texttt{int keys[n]}, 
\texttt{first{\_}pts[n]}, \texttt{flag[n]}.

\begin{figure}[h!]
	\centering
	\subfigure[Test for points derived from mesh models in Mode 1]{
		\label{fig:Subprocedure_K20c}       
		\includegraphics[scale = 0.5]{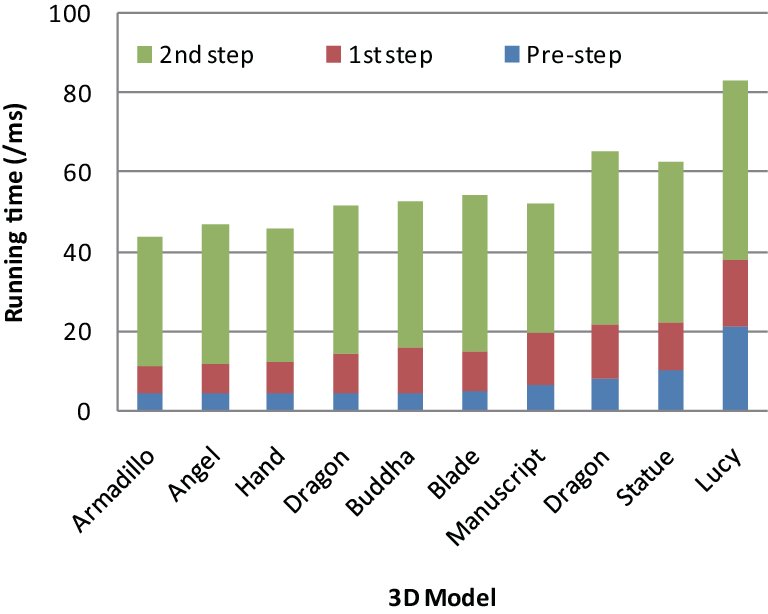}
	}
	\hspace{1em}
	\subfigure[Test for points derived from mesh models in Mode 2]{
		\label{fig:Subprocedure_K20d}       
		\includegraphics[scale = 0.5]{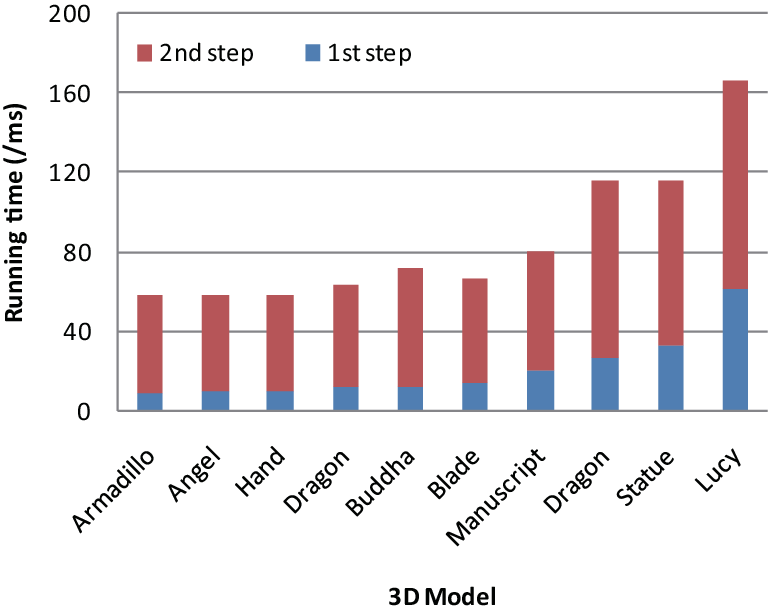}
	}
	\caption{Computational efficiency of sub-procedures on K20c}
	\label{fig:subprocedure}       
\end{figure}

\subsection{Performance of Each sub-procedure}

There are three main sub-procedures in our implementation when adopting the 
preprocessing, i.e., the preprocessing procedure (pre-step), the splitting 
of points into two subsets (1st step), and the recursive procedure of 
finding the expected convex hull (2nd step). To find the potential 
performance bottleneck, we have investigated the computational efficiency of 
these three sub-procedures; see the computational efficiency on K20c presented in Figure \ref{fig:subprocedure}. We have found that in most cases: 
(1) the most computationally expensive step is the 2nd step; (2) the 
most computationally inexpensive one is the pre-step. 
Therefore, the potential performance bottleneck of our implementation is 
probably the 2nd step, and needs to be optimized in further work.

\section{Conclusion}
\label{sec:conclusion}

We have presented a novel implementation of the two-dimensional QuickHull 
algorithm on the GPU. In our implementation, we have transformed the 
Divide-and-Conquer procedures into the operations for segments directly in 
the input arrays. We have strongly utilized several efficient primitives 
including parallel sort, scan, segmented scan, and partitioning provided by 
the library \textit{Thrust}. We have also evaluated 
the performance of our implementation using: (1) points randomly 
distributed in unit square and (2) points derived from 3D mesh 
models with or without employing 
the preprocessing procedure. We have found that our implementation can 
achieve a speedup of up to 10.98x over the Qhull library. We have also observed that it cost about 0.2s to 
find the convex hull of 20M points. We hope that our work can help 
develop efficient implementations of other 
Divide-and-Conquer algorithms on the GPU.




\textbf{Acknowledgments}
The authors are grateful to the anonymous referee 
for helpful comments that improved this paper. This research was supported 
by the Natural Science Foundation of China (Grant Nos. 40602037 and 
40872183).

%
\label{sect:bib}
\bibliographystyle{plain}
\bibliography{ref}


\end{document}